\documentclass{PoS}

\title{Hybrid detection of high-energy cosmic neutrinos with the next-generation neutrino detectors at the South Pole}

\ShortTitle{Hybrid detection of UHE neutrinos}

\author{\speaker{S. Toscano}$^{a}$,
P. Coppin$^{b}$,
K. D. de Vries$^{b}$,
N. van Eijndhoven$^{b}$, 
J. A. Aguilar$^{a}$
\\
\llap{$^{a}$} Universit\'e Libre de Bruxelles, Science Faculty CP230, B-1050 Brussels, Belgium \\
\llap{$^b$} Vrije Universiteit Brussel, Dienst ELEM, B-1050 Brussels, Belgium \\
E-mail:  \email{toscano@icecube.wisc.edu}}

\abstract{In 2013 the IceCube collaboration announced the discovery of a cosmic neutrino flux up to PeV energies, validating neutrino astronomy as the next promising observational technique to explore the high-energy Universe. The neutrino community is moving forward with the construction of new facilities to enhance the detection of these elusive particles at higher energies (up to and beyond EeV) and to increase the statistics at the high-energy end of the IceCube neutrino flux.
Future large volume neutrino detectors, using both the radio Askaryan and the optical Cherenkov signal, will open the possibility of hybrid detection of neutrino interactions within the polar ice.
In this contribution we present a first calculation of the expected number of events for a simplified geometry of one radio station located at 200 m depth in the vicinity of a $\sim$ 10 km$^3$ in-ice Cherenkov detector, similar to the planned IceCube-Gen2 neutrino observatory.
Preliminary simulations show that a total event rate of $\sim 1$ event/year is achievable for a 10-stations array assuming that the Askaryan radio detectors can lower their energy threshold down to $\sim$ PeV energies. Such a possibility is currently under study for the future radio extension foreseen as one of the surface components of IceCube-Gen2.}

\FullConference{36th International Cosmic Ray Conference -ICRC2019-\\
		July 24th - August 1st, 2019\\
		Madison, WI, U.S.A.}

\begin{document}

\section{Introduction}
The recent discovery of cosmic high-energy neutrinos by IceCube \cite{IC2014, ICScience} has paved the path to multi-messenger astronomy, heralding the beginning of an all new observational era of the extragalactic non-thermal Universe.
Cosmic high-energy neutrinos ($E > 100$~TeV) are produced in violent phenomena when for instance cosmic-ray protons are accelerated up to PeV energies and beyond, and interact with surrounding photons and matter. Being weakly interacting particles, neutrinos can escape unscathed from the inner part of cosmic accelerators, providing an unobstructed and intact view of their sources. Astrophysical neutrinos reaching the Earth can be detected capturing the faint Cherenkov emission from the secondary charged particles produced in their interactions with a transparent medium, such as water or the Antarctic ice, as it is done in IceCube.

At energies above 100 PeV, the cosmogenic production dominates the cosmic neutrino flux. When ultra high-energy cosmic rays (UHECRs, $E>10^{18}$ eV) interact with the cosmic microwave background (CMB) or extragalactic background light (EBL) photons via the $\Delta$ resonance, they produce ultra high-energy (UHE) neutrinos from the decay of the resulting pions (an effect known in literature as the Greisen-Zatsepin-Kuzmin suppression of the cosmic ray flux at Earth \cite{GZK}). These cosmogenic, or GZK, neutrinos represent a diffuse flux and can probe the nature of the enigmatic sources of cosmic rays by providing information about the mass composition of the UHECRs. The detection of cosmogenic neutrinos will solve the question of whether the observed high-energy cut-off in the cosmic-ray energy spectrum is indeed due to the GZK effect or due to the exhaustion of the acceleration mechanism at the sources \cite{Auger, TA}. In this context, even a null observation of GZK neutrinos will help to shed light on the existing picture. 

Due to the steeply falling flux, the detection of cosmogenic neutrinos is challenging and requires the instrumentation of Teraton-scale volumes. Despite the remarkable success of the optical Cherenkov technique, the IceCube design is not scalable to such large volumes, and a different technique is needed to detect neutrinos above 100 PeV. Neutrino interactions in dense media produce a coherent, impulsive radio signal (Askaryan effect \cite{Askaryan, AskaryanIce}) that can propagate in transparent media such as the Antarctic ice. Several neutrino experiments (ARA\cite{ARA}, ARIANNA\cite{ARIANNA}, ANITA\cite{ANITA}) are currently exploiting the high transparency of the Antarctic ice at radio frequencies, where UHE neutrino signatures are visible over distances of the order of 1 km. \\

\section{Hybrid detection channel}
\begin{figure}
     \includegraphics[width=1.\textwidth]{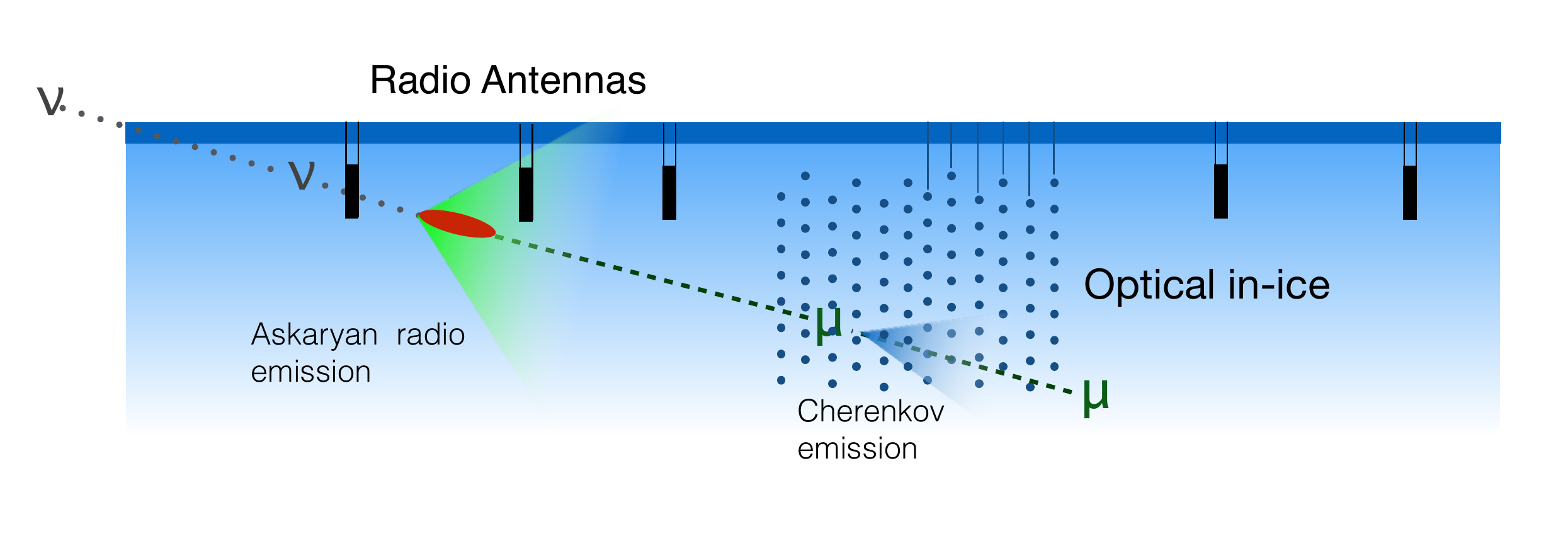}
     \caption{Sketch of the detection principle for a hybrid event. The Askaryan emission (in green) generated at the interaction point is detected by the radio array. The resulting lepton (a muon in this case) propagates through the ice generating Cherenkov photons (in blue) detected by the in-ice detector.}
     \label{figHybrid}
\end{figure}

Driven by IceCube's breakthrough, the IceCube collaboration is proposing a wide-band neutrino facility covering energies from 10 TeV to beyond 100 EeV, dubbed IceCube-Gen2, to be built within the next decade. The future observatory will include an in-ice optical array (with an instrumented volume of $\sim 10 ~\mathrm{km}^3$) for detailed studies of the astrophysical neutrino flux, and a radio array (200 stations covering an area of $\sim ~500 ~\mathrm{km}^2$ ) for detecting cosmogenic neutrinos \cite{Gen2, Gen2WP}.\\
The vicinity of the radio and optical detectors opens the possibility to investigate a new detection channel, the hybrid detection \cite{HybridORA}.
Charged current (CC) interactions of $\nu_\mu$ or $\nu_\tau$ events produce a particle cascade and a long-ranged charged lepton. The coherent Askaryan radio signal generated by the particle showers in the ice is detectable by broad-band radio antennas. At the same time the resulting high-energy lepton can traverse and be detected by the in-ice optical detector. A sketch of the detection principle is shown in Fig.\ref{figHybrid}. 

Similarly, all-flavor neutrino neutral current interactions (NC) and $\nu_e$ CC interactions could produce a high-energy cascade inside the in-ice optical detector also detectable by the radio array. This combined detection will provide an excellent handle on both energy as well as direction and flavor identification of the original neutrino. Moreover, it will help to lower the energy threshold of the radio detection using the optical in-ice detector to trigger the radio detection. 
To study the viability of this new detection channel we have made a first calculation of the expected number of events from muon neutrinos. We search for coincident events in a single radio station together with the in-ice optical detector based only on geometrical effects and energy threshold considerations.

\section{Simulation of high-energy neutrino interactions}
\label{sec:NuGen}
To characterize the performance of the hybrid detection technique, we simulated the interactions of high-energy astrophysical neutrinos in the polar ice sheet. An isotropic astrophysical neutrino flux
\begin{equation}
  E^2\phi(E)=0.9\cdot 10^{-8}\left(\frac{E}{100\ \mathrm{TeV}}\right)^{-0.13}\mathrm{GeV\ cm^{-2}\ s^{-1}\ sr^{-1}}\ ,
 \label{eq:flux}
\end{equation}
is assumed \cite{ICflux}, where the normalization is per flavor and includes both neutrinos and anti-neutrinos. At energies exceeding 1 PeV, the probability that neutrinos arrive from below the horizon is greatly suppressed due to Earth absorption. We therefore only consider neutrinos whose zenith angle is less than $90^\circ$.\par
The medium for the neutrinos to interact with is the 3 km-thick ice sheet at the geographical South Pole. Neutrinos and anti-neutrinos of all flavors can interact with the ice nucleons via CC and NC interactions. In addition, anti-electron neutrinos can interact with electrons via the Glashow resonance. This channel dominates the interaction cross section around the resonance energy of $m^2_W/2m_e\approx 6.3$ PeV. While the cross section for the Glashow resonance can be calculated analytically, this is not the case for CC and NC interactions. As they depend on the parton distribution functions, we use the CTEQ5 model \cite{CTEQ5} to numerically evaluate their cross sections.\par
Different interaction types will lead to the production of different secondary particles. NC interactions always lead to a hadronic cascade with energy $y\cdot E_\nu$, where $E_\nu$ is the initial energy of the neutrino and $y\in [0,1]$ is the inelasticity parameter. Hadronic cascades are also present in CC interactions, though they are accompanied by an energetic electron, muon or tauon, which carries the remaining energy $(1-y)E_\nu$. Glashow events result in the on shell production of a W-boson, which then immediately decays into hadrons (67.4\%) or leptons (32.6\%) \cite{PDG}. In the case of the lepton decay channel, the event topology will depend on the flavor of the lepton. Electrons induce an electromagnetic cascade. Muons, in contrast, can travel several kilometers through the ice, stochastically losing energy as they go along. Tauons lead to more complex event topologies, as they will decay after an average distance $50 \cdot E_\tau / (1\;\mathrm{PeV})$ meters, producing a second cascade. In this work, we consider all interactions that involve a single cascade and a muon, providing a conservative estimate of the expected total rate of neutrino interactions.\par

\section{Hybrid detection simulation}
\label{sec:simu}
The hybrid detector in our study consists of one point, representing the radio antenna at (X,Y,Z) =  (0, 0, -200 m), and a cylinder (height=1~km, radius=1.5~km) representing the in-ice detector located within 1500~m and 2500~m below the surface. The antenna is positioned at a distance $d = 1$ km from the edge of the in-ice optical detector footprint.\par 

The simulation chain consists of the following steps: 
\begin{itemize}
    \item A sample of $2\times10^{6}$ neutrino interactions of all flavors (including Glashow) are generated and uniformly distributed in solid angle. CC and NC interactions are generated with an energy spectrum of $E^{-1.1}$ with $E = [10^{6} -10^{11}]$ GeV. Every event has been given a weight (expressed in unit of [yr$^{-1}$]) used to renormalize the generated spectrum to the astrophysical flux. The generation volume is a cylinder with height = ice sheet ($\sim 3~\mathrm{km}$) and radius  = 4~km\footnote{Although the attenuation length for radio wavelength is of $\sim$ 1 km, radio signals can reach the antenna from a longer distance, within a radius of $\sim$ 4 km, depending on the energy of the particle shower.}. For this study we select only $\nu_\mu$ CC interactions. 
    \item A threshold effect for the radio detection is applied by selecting only hadronic showers generated in $\nu_\mu$ CC interaction with $E > E_{th}$. The currently operating Askaryan detectors have a trigger threshold above 100 PeV. New interferometric techniques have been successfully implemented within the ARA experiment and can be used to lower the energy threshold down to 10 PeV \cite{PA1, PA2}. Moreover, the development of a hybrid trigger, where the information of a detected event can be passed from the optical to the radio detector, opens up the possibility to lower the trigger threshold down to 1 PeV. We study the case for $E_{th}$ = [1 PeV, 10 PeV].
    \item The geometry of the Askaryan emission is taken into account, selecting only events visible by the radio antennas. This coherent impulsive radio Cherenkov radiation is generated from the charge asymmetry in the particle showers at the interaction point. Due to coherence effects, the strongest emission comes at angles close to the radio     Cherenkov angle ($\sim 56^\circ$). Only particle showers having a direction within the Cherenkov cone ($\theta_c \pm 2.5^\circ$) are selected. 
    \item Accompanying muons intersecting the in-ice optical cylinder volume are selected. The energy loss of the muon is calculated analytically using Eq.~\ref{eq:dEdx}:
    \begin{equation}
        \frac{dE}{dx} = a + b \cdot E,
        \label{eq:dEdx}
    \end{equation}
    where a = 0.259 [GeV/mwe], b = 0.363$\cdot10^{-3}$ [mwe$^{-1}$] \cite{dEdx}. From Eq.~\ref{eq:dEdx} we can derive the energy at the entrance of the in-ice volume: 
    \begin{equation}
        E_\mu = \frac{1}{b} \cdot \Big(\frac{a + b \cdot E}{e^{b\cdot l_{track}}}-a\Big),
        \label{eq:Ef}
    \end{equation}
    where $l_{track}$ is the muon track lenght in mwe.
    An additional cut ($E_\mu > 10$ TeV) selects only those muons leaving a detectable signal in the optical Cherenkov detector.
\end{itemize}

\section{Effective area, sensitivity and expected event rate}

To calculate the expected number of events we express the detection efficiency in terms of neutrino effective area. The effective area is calculated using Eq.~\ref{eq:Aeff}:
\begin{equation}
    A_{eff} (E) = N(E) \cdot \frac{\sum_{i}w_i}{\Delta\Omega\cdot\phi(E)}\ \mathrm{cm}^2\ , 
    \label{eq:Aeff}
\end{equation}
where $N(E)$ is the number of events for a given energy $E$, $w_i$ is the event weight as defined in Sec.~\ref{sec:simu}, $\phi$ is the astrophysical neutrino flux as defined in Eq.~\ref{eq:flux}, and $\Delta\Omega$ is the solid angle.

\begin{figure}
\centering
\includegraphics[width=.5\textwidth]{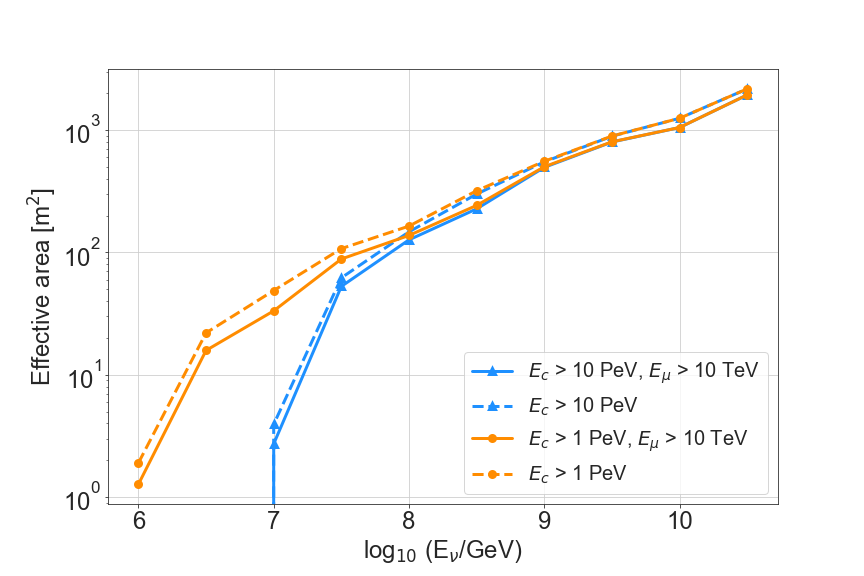}\includegraphics[width=.5\textwidth]{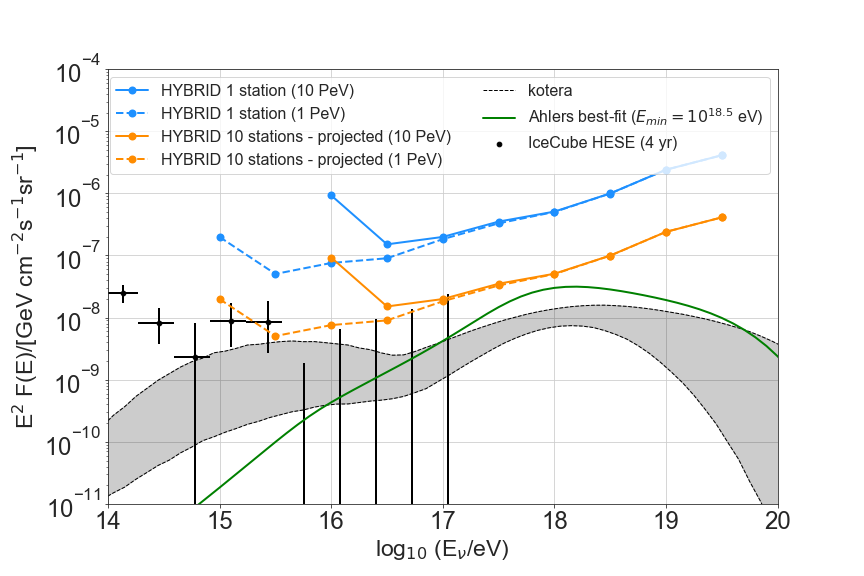}
     \caption{\emph{Left}: Effective area as a function of the energy for different cascade energy thresholds. The dashed lines are obtained considering all muons reaching the in-ice detection volume, while solid lines include only muons with an energy $E_\mu$> 10~TeV at the entrance of the in-ice detector. \emph{Right}: Hybrid detector sensitivity for one antenna at 200~m depth and an in-ice volume similar to what is planned for IceCube-Gen2. The detector sensitivity is compared with measurements of the astrophysical flux from IceCube \cite{IC2014}, and the most reasonable GZK models taken from Kotera et al.~\cite{Kotera} and Ahlers et al.~\cite{Ahlers}. The projection for a hybrid layout with 10 stations is also shown.}
     \label{figSensitivity}
\end{figure}

The effective area is shown in the left plot of Fig.~\ref{figSensitivity} as a function of the neutrino energy for two different energy thresholds (1 PeV and 10 PeV) of the radio detection (dashed lines), and a muon energy ($E_\mu$) above 10~TeV at the entrance of the in-ice optical volume (solid lines). 

Given the effective area, the detector sensitivity to an isotropic neutrino flux for a zero background experiment is calculated using Eq.~\ref{eq:sensitivity}:
\begin{equation}
    S(E) = \frac{1}{4\pi\cdot A_{eff}\cdot T} .
    \label{eq:sensitivity}
\end{equation}

Fig.~\ref{figSensitivity} shows the hybrid sensitivity for $E_{th}$ = [1~PeV, 10~PeV] and $E_\mu$ > 10 TeV together with the astrophysical neutrino flux measured by IceCube \cite{IC2014} and two GZK neutrino models taken from \cite{Kotera, Ahlers}. The projected sensitivity for a hybrid detector with 10 antennas/stations is shown for comparison.

\begin{figure}
     \includegraphics[width=.5\textwidth]{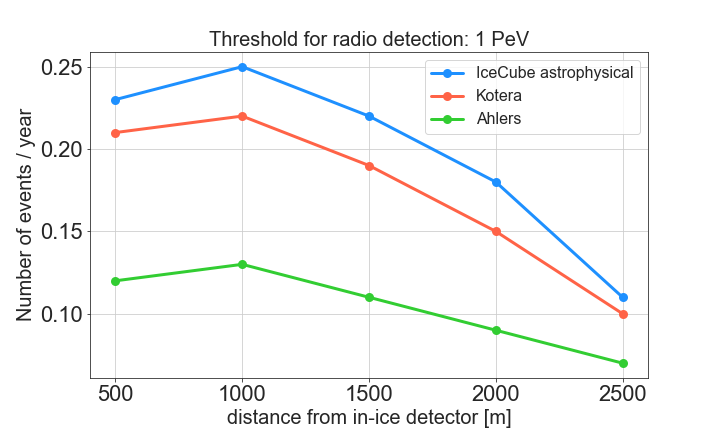}\includegraphics[width=.5\textwidth]{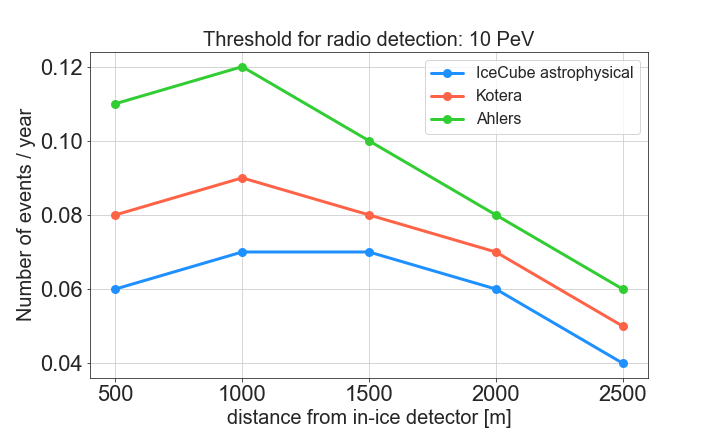}
     \caption{Number of expected hybrid events per year as a function of the radio-antenna distance from the in-ice detector, for 1 PeV threshold (left) and 10 PeV threshold (right) of the radio detection. No condition is imposed to the muon energy.}
     \label{figRatevsDistance}
\end{figure}

The number of expected events can be calculated directly from the effective area using Eq.~\ref{eq:rate}:
\begin{equation}
    N_\nu = 4\pi\cdot\int{dE\cdot A_{eff}\cdot \frac{dN}{dE}},
    \label{eq:rate}
\end{equation}
where $dN/dE$ represents the differential neutrino flux from different model predictions.\\
Fig.~\ref{figRatevsDistance} shows the event rate (per year) as a function of the antenna distance from the in-ice optical detector. The two plots show results for the two thresholds at 1 PeV (left) and 10 PeV (right) for the two GZK models considered for this study and the expectation from the IceCube astrophysical flux, assuming that no cut-off is observed at high energies. We find an optimal distance of $\sim$ 1~km. As expected, the number of detectable events strongly depends on the trigger threshold of the radio detector. In particular, the number of events expected from the astrophysical flux decreases considerably when a threshold of 10 PeV is applied. This is also visible from the sensitivity plot shown in Fig.~\ref{figSensitivity}.    

\begin{figure}
\centering
     \includegraphics[width=.6\textwidth]{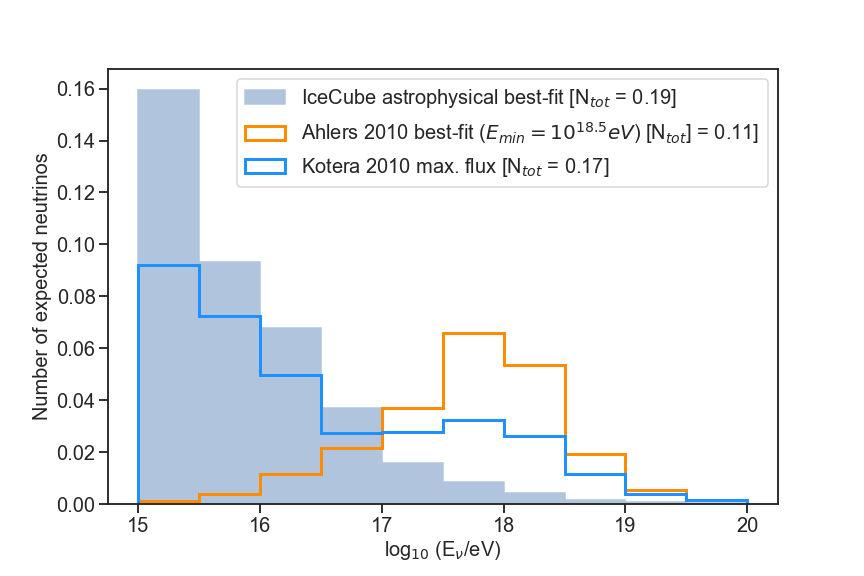}
     \caption{Energy distribution of the hybrid events expected from different GZK models and the IceCube astrophysical flux. The histograms show the case for a radio detection threshold of 1 PeV and a final muon energy reaching the in-ice detector bigger than 10 TeV. The total number of events per year (integral of the histograms) is reported in the legend.}
     \label{figRateHisto}
\end{figure}

Considering the best case scenario of a threshold at 1 PeV, and imposing $E_\mu$ > 10~TeV, we calculate the energy distribution of detectable events in one year shown in Fig.~\ref{figRateHisto}. 
The total number of expected neutrinos varies from 0.1 to 0.2 events/year depending on the considered flux. To maximise the effective volume, radio stations are generally deployed at a distance of 1.5-2~ km, so that they will monitor independent volumes. For this reason, the expected number of events increases linearly with the number of stations. For a hybrid detector layout with 10 antennas an event rate $\sim1-2$ events/year is achievable.  

\section{Conclusions}

Plans to build a new neutrino facility within the next decade have already started at the South Pole. The IceCube collaboration is proposing the construction of the next-generation neutrino observatory IceCube-Gen2. The future observatory will include a high-energy optical in-ice array extension of the IceCube detector, and a radio array for UHE neutrino detection. 
In this contribution we have studied the viability of a radio-optical coincident (hybrid) detection, considering only one radio antenna and an in-ice volume similar in size to what is planned for IceCube-Gen2. The event rate expected from the hybrid channel depends on the trigger threshold of the radio detector. For a threshold of 1 PeV, a rate of $\sim1-2$ events/year is expected for an optimized layout with about 10 radio stations. The expected event rate for a higher threshold of 10 PeV, achievable by the current instruments, remains too small even for an extended hybrid configuration with 10 antennas. The feasibility to lower the trigger threshold to 1 PeV needs to be investigated further. One possible avenue is the design of an \emph{ad-hoc} hybrid trigger, where a trigger is sent from the in-ice optical to the radio detector. Although challenging, the possibility of such trigger could be investigated within the proposed Radio Neutrino Observatory (RNO)~\cite{RNO}, a pathfinder for the Gen2-radio component. Alternatively, a new radio detection technique based on radar reflection has been recently proposed \cite{RADAR}. High-energy neutrinos interacting in ice will induce a particle shower that, moving at the speed of light, will ionize the medium leaving behind a plasma tube made of electrons and protons. First studies show that a radar device will be able to measure the reflection of radio waves from the induced ionization plasma \cite{RADAR2}, opening the possibility to cover the unexplored energy region between several PeV, where IceCube runs out of events, and a few EeV, where the Askaryan detectors begin to have large effective volumes. The lower energy threshold of the radar detection with respect to the direct Askaryan radio signal, and the $4\pi$ distribution of the scattered signal will boost the hybrid technique in a cost-effective way.

\section*{Acknowledgment}

The authors acknowledge the funding from the Belgian Funds for Scientific Research (FRS-FNRS and FWO), the FWO programme for International Research Infrastructure (IRI), and ERC-StG (no. 805486, K.D. de Vries) of the European Research Council (ERC).

\end{document}